\documentclass{PoS}

\usepackage[utf8]{inputenc}
\usepackage{graphicx}
\usepackage{xcolor}
\usepackage{amsmath}
\usepackage{bm}

\newcommand{\V}{n}

\newcommand{\ed}{{e}}

\newcommand{\peq}{p}

\newcommand{\beshydro}{{\sc BEShydro}}

\newcommand{\eq}{{\,=\,}}
\newcommand{\snn}{\sqrt{s_\mathrm{NN}}}
\newcommand{\etas}{$\eta_s$}
\newcommand{\tauI}{\tau_{i}}

\graphicspath{{fig/}}

\title{Baryon diffusion near the QCD critical point}

\ShortTitle{Baryon diffusion near the QCD critical point}

\author{\speaker{Lipei Du}\thanks{Supported in part by the U.S. Department of Energy (DOE), Office of Science, Office for Nuclear Physics, under Award No.~\rm{DE-SC0004286} and within the framework of the BEST Collaboration, and in part by the National Science Foundation (NSF) within the framework of the JETSCAPE Collaboration under Award No.~\rm{ACI-1550223}.}\\
        Department of Physics, The Ohio State University, Columbus, Ohio 43210, USA\\
        E-mail: \email{du.458@osu.edu}}

\author{Xin An\\
        Department of Physics and Astronomy, University of North Carolina, Chapel Hill, North Carolina 27599, USA\\
        E-mail: \email{xan@unc.edu}}
        
\author{Ulrich Heinz\\
        Department of Physics, The Ohio State University, Columbus, Ohio 43210, USA\\
        E-mail: \email{heinz.9@osu.edu}}     
        
\abstract{
    Fireballs created in relativistic heavy-ion collisions at different beam energies have been argued to follow different trajectories in the QCD phase diagram in which the QCD critical point serves as a landmark. Using a (1+1)-dimensional model setting with transverse homogeneity, we study the complexities introduced by the fact that the evolution history of each fireball cannot be characterized by a single trajectory but rather covers an entire swath of the phase diagram, with the finally emitted hadron spectra integrating over contributions from many different trajectories. Studying the phase diagram trajectories of fluid cells at different space-time rapidities, we explore how baryon diffusion shuffles them around, and how they are affected by critical dynamics near the QCD critical point. We find a striking insensitivity of baryon diffusion to critical effects. Its origins are analyzed and possible implications discussed.
}

\FullConference{CPOD2021 - International Conference on Critical Point and Onset of Deconfinement\\
                March 15 - March 19, 2021\\
                Online}

\begin{document}

\section{Introduction}

Confirming the existence and finding the location of the hypothetical critical point (CP) \cite{Stephanov:2006zvm} in the phase diagram of Quantum Chromodynamics (QCD) in principle can be achieved by heavy-ion collisions which have been carried out at different experimental facilities and at various beam energies \cite{Bzdak:2019pkr}. One of the most promising signatures of the QCD CP, based on static equilibrium considerations, is a non-monotonic beam energy dependence of higher-order cumulants of the fluctuations in the net proton production yields \cite{Stephanov:2008qz}. -- Experimental measurements hinting at such a non-monotonicity were reported recently by the STAR Collaboration  \cite{STAR:2020tga}.

Unfortunately, the fireballs created in heavy-ion collisions are highly dynamical whose rapid expansion keeps the thermodynamic environment and the critical fluctuations out of equilibrium. Thus, to confirm or exclude the CP via systematic model-data comparison, reliable dynamical simulations of off-equilibrium critical fluctuations and the associated final particle cumulants, on top of a well-constrained comprehensive dynamical description of the bulk medium at various beam energies, are indispensable \cite{An:2021wof}. For this purpose the Hydro+/++ framework \cite{Stephanov:2017ghc,An:2019csj} incorporating off-equilibrium critical fluctuations was developed, but a comprehensive multi-stage and fully validated framework for heavy-ion collisions at low Beam Energy Scan (BES) energies is still missing \cite{An:2021wof}. 

The situation is made even more complicated by the back-reaction of the critical fluctuations on the bulk evolution of the medium. Critical effects on the bulk viscous pressure were shown to have non-negligible phenomenological consequences on the rapidity distributions of hadronic particle yields \cite{Monnai:2016kud}, implying that critical effects might indeed play an important role in the calibration of the bulk medium. To gain further guidance on how to deal with the critical effects when constraining the bulk dynamics, we study here critical effects on the bulk evolution through baryon diffusion \cite{Du:2021zqz}. We also explore different viscous effects on the phase diagram trajectories along various space-time rapidities of the fireball.

\section{Criticality of baryon diffusion}\label{sec:hydro}

In the hydrodynamic description of heavy-ion collisions the conservation equations for energy-momentum and net baryon charge 
are formulated covariantly as \cite{Denicol:2018wdp, Du:2019obx}
\begin{equation}\label{eq:conservation}
d_\mu T^{\mu\nu} = d_\mu(\ed u^{\mu}u^{\nu}-\peq\Delta^{\mu\nu}-\Pi\Delta^{\mu\nu}+\pi^{\mu\nu}) =  0\,, \quad d_\mu N^\mu = d_\mu(n u^{\mu}+ n^{\mu}) = 0\,.
\end{equation}
Here $d_\mu$ is the covariant derivative in Milne coordinates, and $T^{\mu\nu}$ and $N^\mu$ are the energy-momentum tensor and (net) baryon current, respectively. $\ed$ and $n$ are the energy density and baryon density in the local rest frame (LRF), $p$ the pressure, $u^\mu$ the four-velocity of the fluid element (LRF chosen as the Landau frame where $u_\mu T^{\mu\nu}=\ed u^\nu,\, u_\mu N^\mu=n$), and $\Delta^{\mu\nu} \equiv g^{\mu\nu} - u^{\mu}u^{\nu}$. The dissipative components including the bulk viscous pressure $\Pi$, the shear stress tensor $\pi^{\mu\nu}$, and the baryon diffusion current $n^\mu$ describe the deviations from local equilibrium.

In this work, to isolate the effects from the baryon diffusion current $n^\mu$, we shall ignore the dissipative effects from $\pi^{\mu\nu}$ and $\Pi$, focusing only on $n^\mu$. The equation of motion for $n^\mu$ from the Denicol-Niemi-Molnar-Rischke (DNMR) theory \cite{Denicol:2018wdp} is an Israel-Stewart type equation:
\begin{equation}
\label{eq:IS_nmu3}
    u^\nu\partial_\nu \V^\mu = -\frac{1}{\tau_{n}}(n^{\mu }-n^{\mu}_{\rm NS})  -\frac{\delta_{nn}}{\tau_n}n^\mu\theta - n^\nu u^\mu D u_\nu
    - u^\alpha \Gamma^\mu_{\alpha\beta} n^\beta\,,
\end{equation}
where $\theta\equiv d\cdot u$ is the scalar expansion rate, $D \equiv u_\mu d^{\mu}$ is the covariant time derivative, and $\Gamma^\mu_{\alpha\beta}$ are the Christoffel symbols. The $n^\mu\theta$-term is the only higher order gradient contribution we keep in this work, and $\delta_{nn}$ is the associated transport coefficient. The baryon diffusion current $n^\mu$ is driven by chemical gradients and relaxes to its Navier-Stokes limit $n^{\mu}_{\rm NS}\equiv\kappa_n \nabla^{\mu} \alpha\equiv\kappa_n \nabla^{\mu}(\mu/T)$ on the scale of its relaxation time $\tau_{n}$, where $\kappa_n$ is the baryon diffusion coefficient, and $\nabla^\mu \equiv \Delta^{\mu\nu}d_\nu$ is the spatial gradient in the LRF. To exhibit all the critical singularities in the Navier-Stokes limit we rewrite it in terms of density and temperature gradients,
\begin{equation}
\label{eq:nmu_NS_decomposition}
    n^{\mu}_{\rm NS} 
    =  \frac{\kappa_n}{T\chi}\nabla^{\mu}n
    +\frac{\kappa_n}{Tn}\left[\left(\frac{\partial p}{\partial T}\right)_n-\frac{\ed+p}{T}\right]\nabla^{\mu}T 
    \equiv D_B\nabla^{\mu}n+D_T\nabla^{\mu}T\,,
\end{equation}
where $\chi\equiv(\partial n/\partial\mu)_T$ is the isothermal susceptibility. The singularity in $\chi$ can be obtained naturally from the Equation of State (EoS) if incorporated properly.

Now we turn to the effects of the CP on baryon transport in a relativistic QCD fluid, belonging to the static universality class of the 3-dimensional Ising model \cite{Berges:1998rc} and the dynamical universality class of Model H in the Hohenberg-Halperin classification \cite{RevModPhys.49.435}. Near the CP, fluctuations at the scale of the correlation length $\xi$ significantly modify the physical thermodynamic and transport coefficients. 
We shall focus on the critical scaling resulting from equilibrium fluctuations for thermodynamic quantities and from analytic non-equilibrium fluctuations for transport coefficients. The second-order thermodynamic quantity $\chi$, as well as the first-order transport coefficient $\kappa_n$, scale with the correlation length as $\chi\sim \xi^2,\, \kappa_n\sim \xi$  \cite{RevModPhys.49.435}, where for simplicity the exponents are rounded to their nearest integers. Therefore, according to Eqs.~\eqref{eq:nmu_NS_decomposition}, $D_B\sim \xi^{-1},\, D_T\sim \xi$. To identify the critical behavior for the relaxation time $\tau_n$, we first note that it characterizes the relaxation time of $n^\mu$ to the Navier-Stokes limit $n^\mu_{\text{NS}}$. Since $n^\mu$ can only equilibrate after {\it all} fluctuating degrees of freedom contributing to $n^\mu$ also equilibrate, $\tau_n$ can be considered as the typical equilibration time scale of the slowest fluctuation mode near the CP. According to Ref.~\cite{An:2019csj}, the {\it slowest} mode contributing to $n^\mu$ is the diffusive-shear correlator between the entropy per baryon density fluctuations $\delta (s/n)$ and the flow fluctuations $\delta u_\mu$: $G\sim\langle\delta (s/n) \delta u_\mu \rangle$.  Near the CP, the relaxation rate for this mode is dominated by contributions with typical wave numbers $q\sim 1/\xi$ and scales as $\Gamma_G \sim \xi^{-2}$. Thus it is natural to expect $\tau_n\sim\tau_G=\Gamma_G^{-1}\sim\xi^2$. Adopting this critical behavior the Israel-Stewart equation \eqref{eq:IS_nmu3} is found to turn into a Hydro+ equation \cite{Du:2021zqz}.

\section{Results and discussion}
\label{sec:results}

In this exploratory study \cite{Du:2021zqz} we focus entirely on the longitudinal dynamics of the baryon diffusion current for Au-Au collisions at $\snn\eq19.6$\,GeV, modeled by a (1+1)-dimensional system without transverse gradients initiated instantaneously at a constant proper time $\tauI\eq1.5$\,fm/$c$, with initial longitudinal profiles taken from Ref.~\cite{Denicol:2018wdp}. We assume a ``static'' initial longitudinal momentum flow profile in Milne coordinates with a vanishing initial baryon diffusion current. For the EoS at non-zero net baryon density we use {\sc neos} \cite{Monnai:2019hkn} from which we obtain the non-critical isothermal susceptibility $\chi_0\equiv(\partial n/\partial\mu)_T$. The non-critical values of the baryon diffusion coefficient ($\kappa_{n,0}$) and the relaxation time ($\tau_{n,0}$) are obtained from kinetic theory \cite{Denicol:2018wdp}. In the critical regime we use the parametrizations
\begin{equation}\label{eq:cri_scaling}
    \chi=\chi_0\left(\xi/\xi_0\right)^2\,,\quad\kappa_n=\kappa_{n,0}\left(\xi/\xi_0\right)\,,\quad\tau_n=\tau_{n,0}\left(\xi/\xi_0\right)^2
\end{equation}
to incorporate their critical scaling. These hold in the entire crossover domain of the QCD phase diagram, both far away from and within the critical region. We use an analytical parametrization of $\xi(\mu, T)$ in which $\xi_\mathrm{max}/\xi_0\eq10$ \cite{Du:2021zqz}. The equations are solved numerically using \beshydro{} \cite{Du:2019obx}.

\subsection{Longitudinal baryon transport}
\label{sec:diffcp}

\begin{figure}[!b]
\begin{center}
\hspace{-0.5cm}
\includegraphics[width=0.35\textwidth]{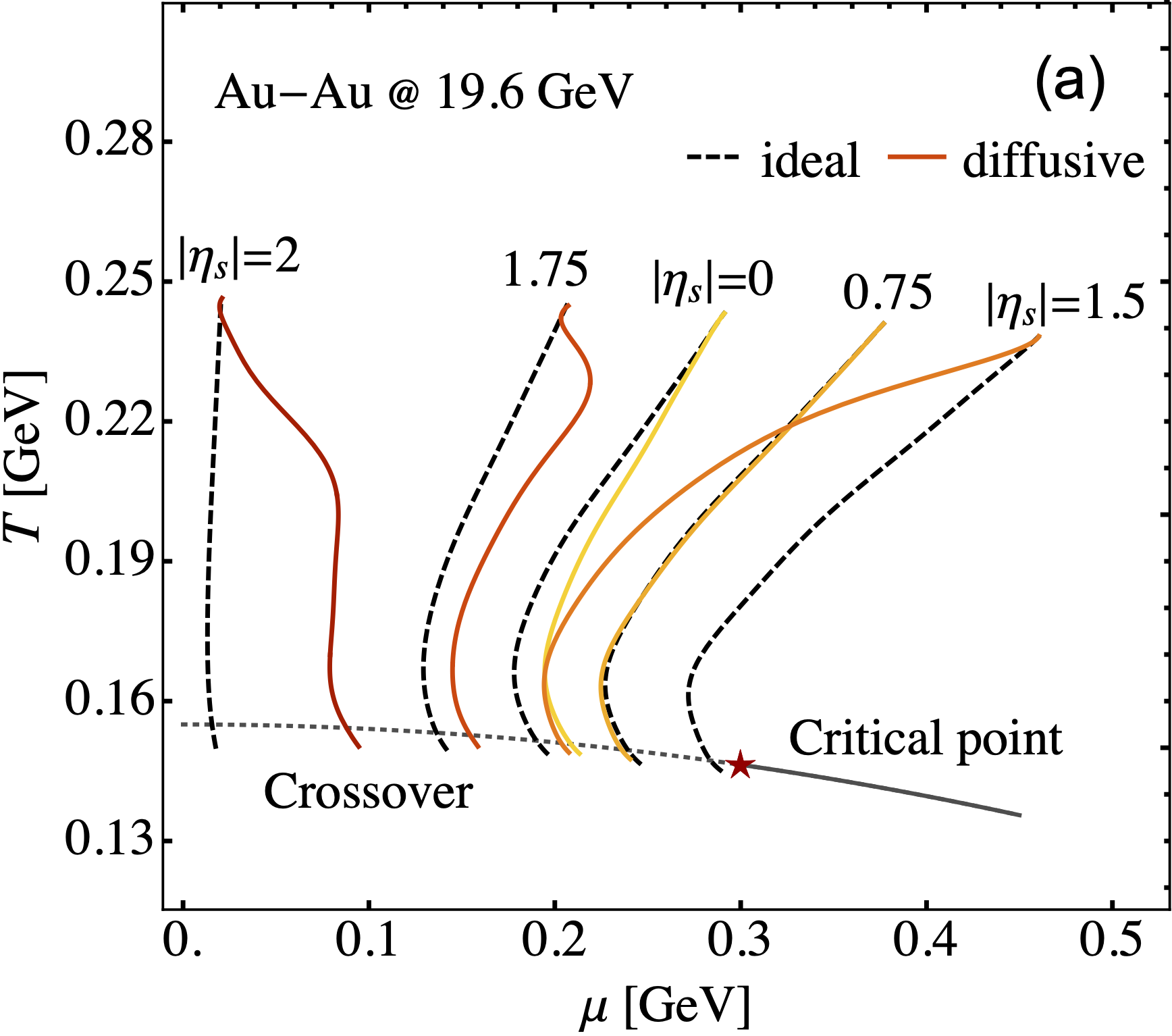}  
\hspace{1.4cm}
\includegraphics[width=0.35\textwidth]{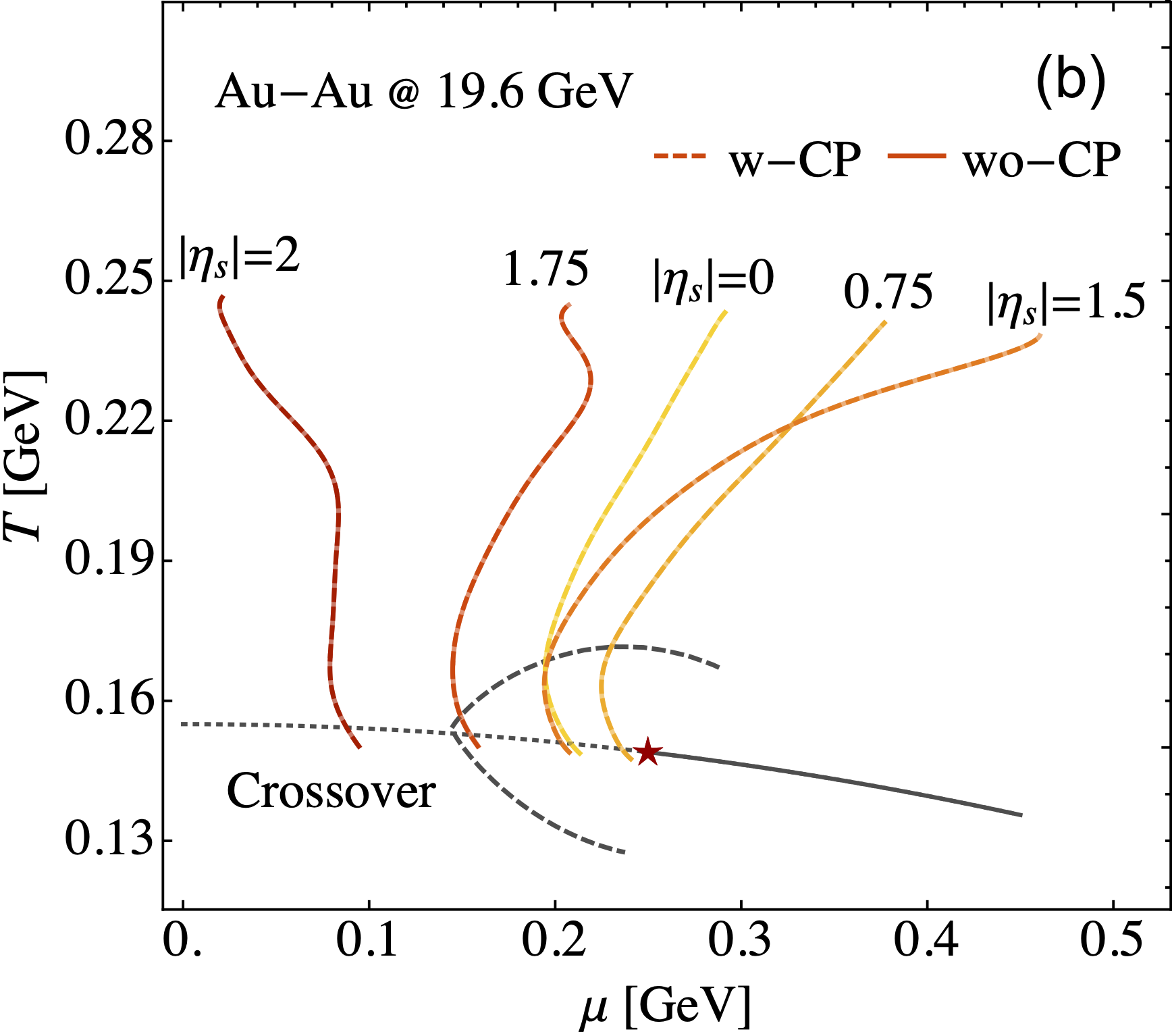}
\caption{%
    Phase diagram trajectories of fluid cells at different $|\eta_s|$ for the Au+Au collision fireball. (a) Black dashed lines indicate ideal evolution while colored solid lines include the effects of baryon diffusion (critical effects not included). The phase transition line and CP are included only to guide the eye. (b) Phase diagram trajectories with (w-CP, colored dashed lines) and without (wo-CP, colored solid lines, same as the ones in panel (a)) inclusion of critical effects. The w-CP case accounts for the critical scaling of all parameters controlling the evolution [Eqs.~\eqref{eq:cri_scaling}]. Figures taken from Ref.~\cite{Du:2021zqz}.
    }
    \vspace*{-10mm}
    \label{fig:phase_dia_traj}
\end{center}
\end{figure}

We first illustrate in Fig.~\ref{fig:phase_dia_traj}a the phase diagram trajectories of fluid cells at several selected $|\eta_s|$ values, both with (diffusive, solid) and without (ideal, dashed) baryon diffusion. The difference between the ideal and diffusive trajectories exhibits a remarkable dependence on \etas{}: Both the sign and the magnitude of the diffusion-induced shift in baryon chemical potential depend strongly on space-time rapidity. In most cases, we note that the diffusive trajectories move initially rapidly away from the corresponding ideal ones, but then quickly settle on a roughly parallel ideal trajectory. The first effect results from strong initial longitudinal baryon transport through baryon diffusion, whereas the second one indicates a fast decay of the diffusion current. Fig.~\ref{fig:phase_dia_traj}a is reminiscent of the QCD phase diagram often shown to motivate the study of heavy-ion collisions at different collision energies in order to explore QCD matter at different baryon doping \cite{Bzdak:2019pkr}. What had been shown there are (isentropic) expansion trajectories for matter created {\em at midrapidity in heavy-ion collisions with different beam energies}; in contrast,  Fig.~\ref{fig:phase_dia_traj}a shows expansion trajectories {\em for different parts of the fireball in a collision with a fixed beam energy}. 

%
%

Fig.~\ref{fig:phase_dia_traj}a makes the point that in general the matter created in heavy-ion collisions can never be characterized by a single trajectory but by a swath of them, and fluid cells at different \etas{} pass through different regions of the QCD phase diagram and therefore are expected to be affected differently by the QCD CP. We study this further by incorporating the critical scaling \eqref{eq:cri_scaling} which indicates that in the proximity of the CP (where $\xi/\xi_0>1$) $\chi$ and $\kappa_n$ are enhanced and thus move the Navier-Stokes target value of $n^\mu$ [see Eq.~\eqref{eq:nmu_NS_decomposition}]. Furthermore, its approach towards the target is critically slowed down since $\tau_n$ increases as $\xi$ grows. Repeating the simulations with the same setup as above, but now including critical scaling, results in the dashed lines shown in Fig.~\ref{fig:phase_dia_traj}b. For the parametrization of the correlation length $\xi(\mu,T)$ we assumed a CP located at ($T_c\eq149$\,MeV,\,$\mu_c\eq250$\,MeV). This is very close to the right-most trajectory which should therefore be most strongly affected by it. Surprisingly, none of the trajectories, not even the one passing the CP in close proximity, are visibly affected by critical scaling of transport coefficients.

%
\begin{figure}[!t]
\begin{center}
\hspace{-0.6cm}
\includegraphics[width=0.41\textwidth]{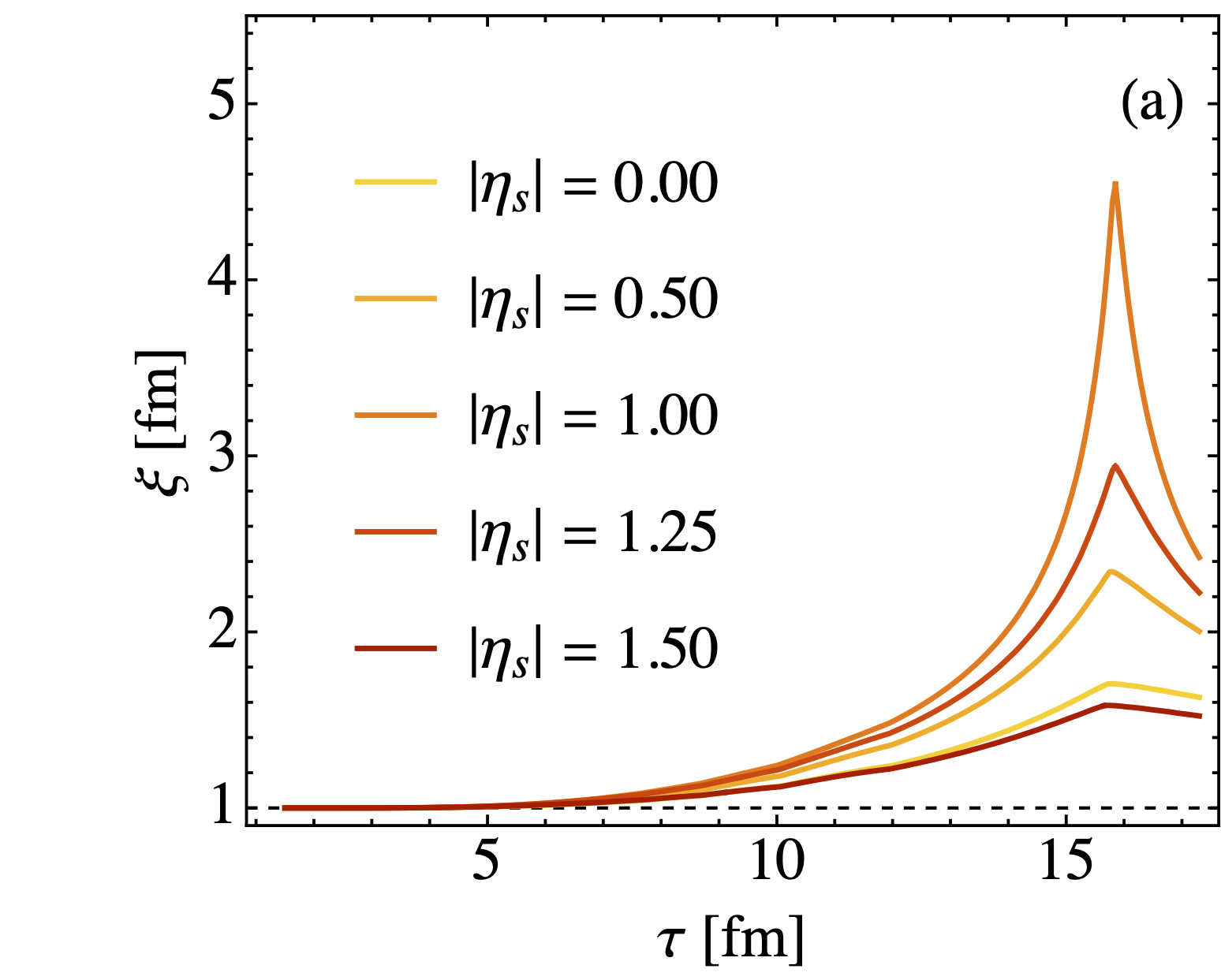}
\hspace{0.8cm}
\includegraphics[width=0.41\textwidth]{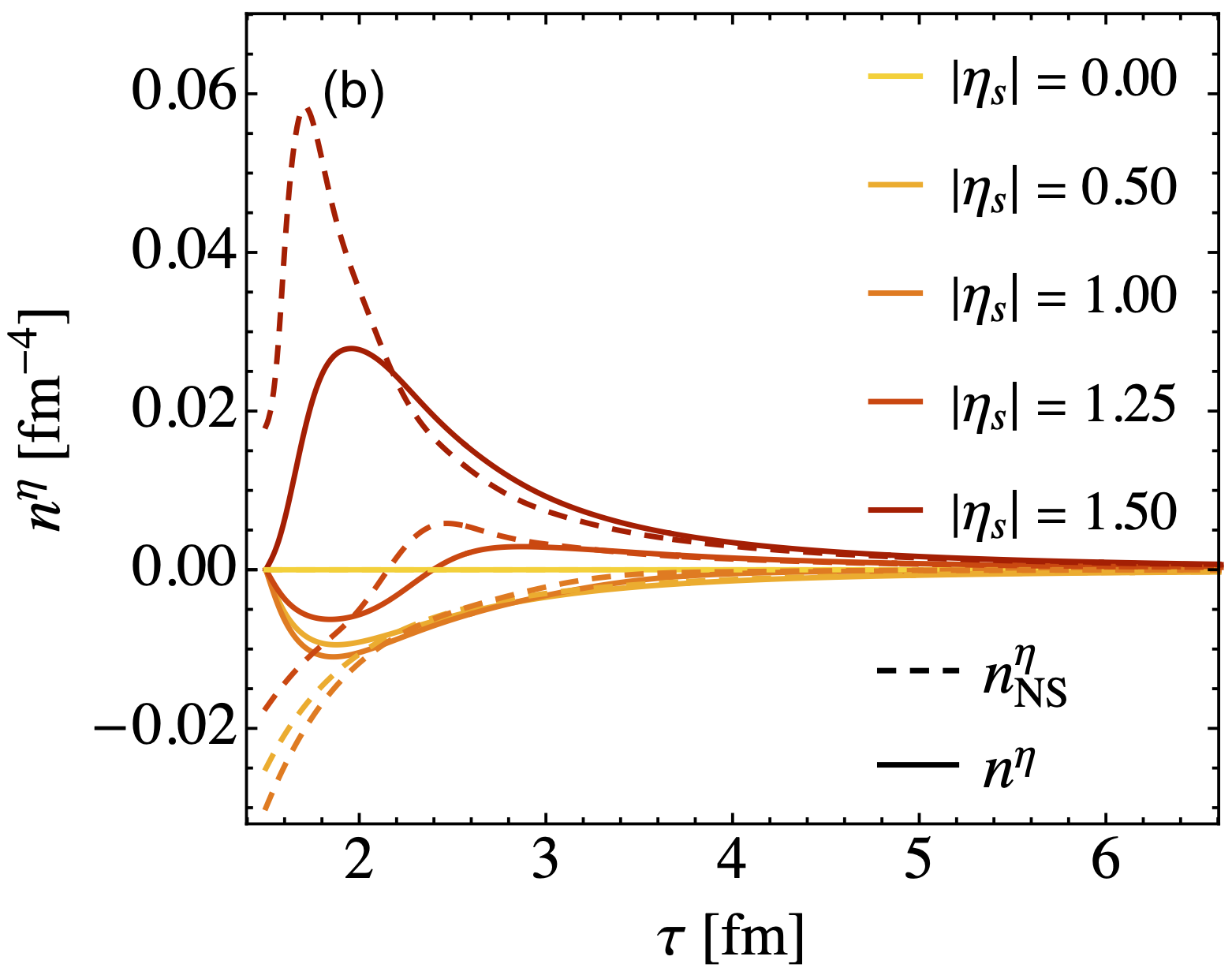}
\caption{%
    Time evolution of (a) correlation length and (b) longitudinal baryon diffusion $n^\eta$ (solid lines) and its corresponding Navier-Stokes limit $n^\eta_\mathrm{NS}$ (dashed lines) at selected space-time rapidities \cite{Du:2021zqz}.
    }
    \label{fig:xi_neta_evolution}
\end{center}
\end{figure}
%

To better understand this we plot in Fig.~\ref{fig:xi_neta_evolution} the history of the correlation length and baryon diffusion current at different $\eta_s$. In Fig.~\ref{fig:xi_neta_evolution}a we see that $\xi$ does show the expected critical enhancement, which, however, does not begin in earnest before the fireball has cooled down to a low temperature just above $T_c$. Fig.~\ref{fig:xi_neta_evolution}b shows that at this late time the baryon diffusion current has already decayed to a tiny value. This two-stage feature, with a first stage characterized by large baryon diffusion effects without critical modifications and a second stage characterized by large critical  fluctuations with negligible baryon diffusion effects on the bulk evolution, is an important observation. Besides, the relaxation time for baryon diffusion increases at late times \cite{Du:2021zqz}, generically as a result of cooling but possibly further enhanced by critical slowing down if the system passes close to the CP, which makes it difficult for the baryon diffusion current to grow again. These facts explain why no sensitivity to the CP was seen in Fig.~\ref{fig:phase_dia_traj}b. On the other hand, the fast decay of $n^\eta$ can be understood through that of its Navier-Stokes limit $n^\eta_\mathrm{NS}$ (dashed lines in Fig.~\ref{fig:xi_neta_evolution}b) since the former has basically relaxed to the latter at $\tau\gtrsim3.5\,$fm$/c$. The decay of $n^\eta_\mathrm{NS}$ has two reasons: (i) The gradients of $\mu/T$ decrease with time, owing to both the overall expansion of the system and the diffusive transport of baryon charge from dense to dilute regions of net baryon density, and (ii) the baryon diffusion coefficient $\kappa_{n,0}$ decreases dramatically, as a result of the fireball's decreasing temperature  \cite{Du:2021zqz}.

\subsection{Viscous effects on phase diagram trajectories}
\label{sec:visctraj}

\begin{figure}[!tb]
\begin{center}
    \hspace{-0.5cm}
    \includegraphics[width= 0.35\textwidth]{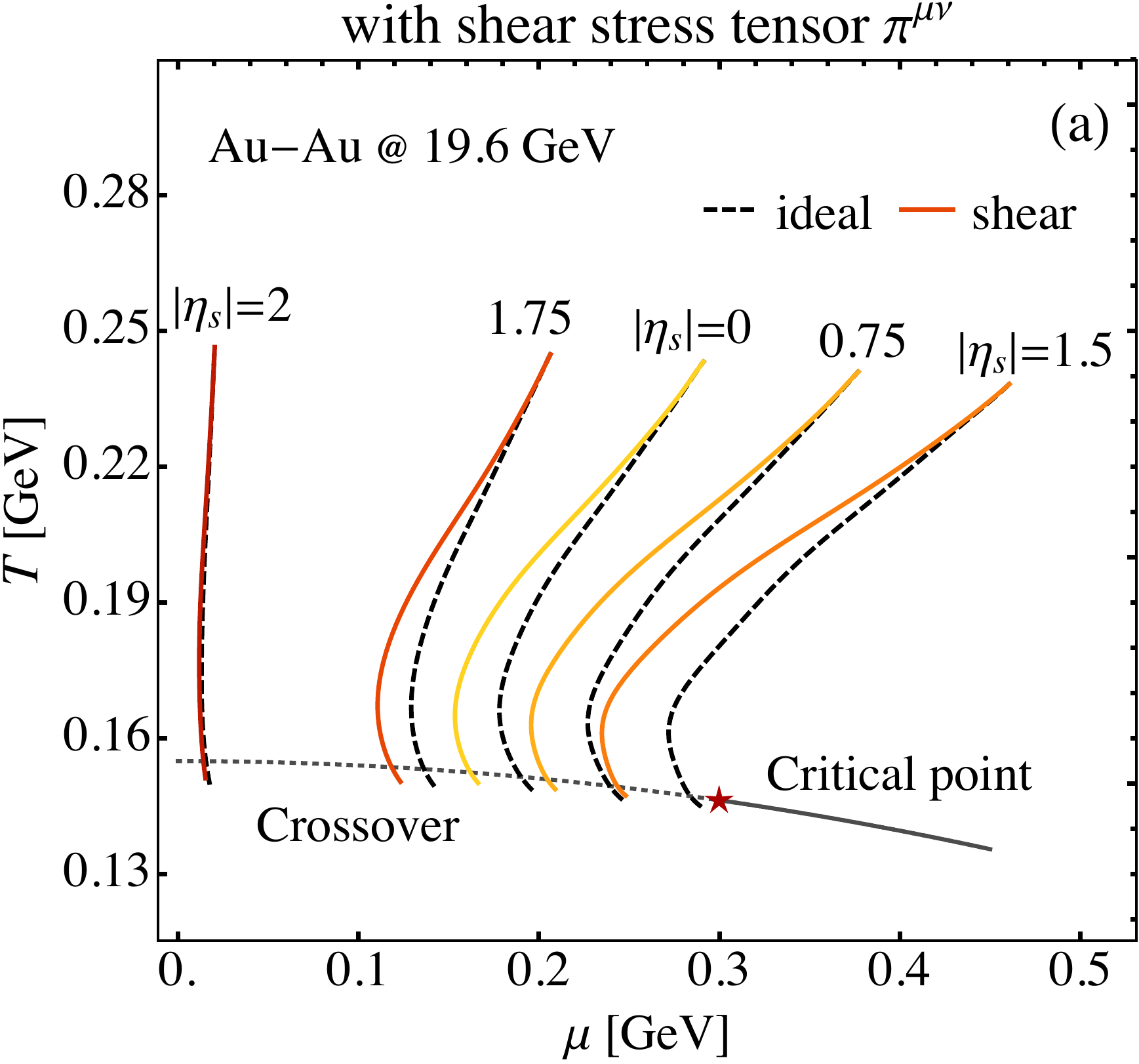}
    \hspace{1.4cm}
    \includegraphics[width= 0.35\textwidth]{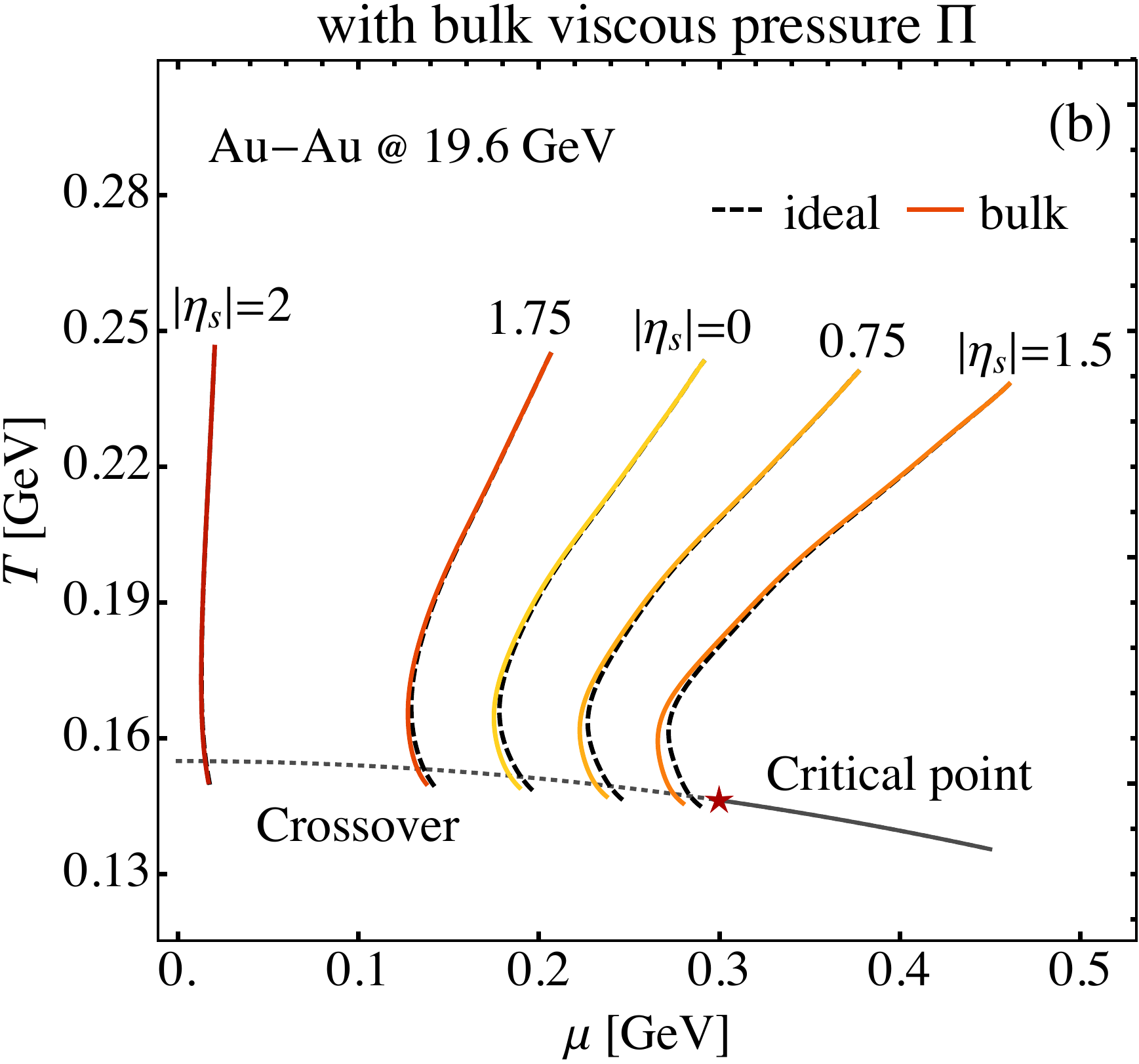}
\caption{Phase diagram trajectories of fluid cells at different space-time rapidities with different viscous effects: (a) shear stress tensor $\pi^{\mu\nu}$ and (b) bulk viscous pressure $\Pi$.}
\label{fig:visctraj}
\end{center}
\end{figure}

We have seen in Fig.~\ref{fig:phase_dia_traj}a that, compared to the ideal case, baryon diffusion reshuffles the phase diagram trajectories at different $|\eta_s|$ (i.e. it introduces crossings) by longitudinally transporting baryon number and thus changing their relative sequence in chemical potential when the system evolves. We emphasize that this feature of baryon diffusion effects originates from the ``interactions'' among trajectories of different fluid cells induced by baryon transport, which also underlies the convergence of $n^\eta$ to a vanishing value at late times shown in Fig.~\ref{fig:xi_neta_evolution}b; in other words, they happen because the diffusion smooths out the longitudinal gradient in $\alpha=\mu/T$. This is different from the previously studied case of trajectories of a single fluid cell undergoing Bjorken expansion with varying initial conditions (e.g.~\cite{Dore:2020jye}). In Fig.~\ref{fig:visctraj} we show analogous effects on the phase diagram trajectories caused by (a) the shear stress tensor $\pi^{\mu\nu}$ and (b) the bulk viscous pressure $\Pi$, separately (easily achievable as \beshydro{} features a modular structure that allows to turn on and off different dissipative components and study their physical effects individually \cite{Du:2021fyr}). We use the same setup as above, setting $\pi^{\mu\nu}=0=\Pi$ initially. We see that, compared to the ideal case, all the trajectories with shear or bulk viscosity are generically pushed to the left or, equivalently, upward, caused by viscous heating. The effect from $\Pi$ is relatively smaller than that of $\pi^{\mu\nu}$, as the bulk viscosity $\zeta/s$ is parametrized to peak at 155\,MeV \cite{Du:2019obx} and hence only plays a role towards the end of the evolution. For different parameters and/or initial conditions shear and bulk stresses could possibly also lead to trajectory crossings, as a result of rapidity-dependent viscous heating effects, but they are unlikely to be as efficient as baryon diffusion which changes their chemical potentials directly. 

\section{Conclusions and discussion}
\label{sec:summary}

In this work we studied a (1+1)-dimensional system without transverse gradients and flow, with initial conditions modeling central Au-Au collisions at $\sqrt{s_\mathrm{NN}} \sim20$\,GeV, to explore diffusive baryon transport along the longitudinal (beam) direction in heavy-ion collisions. We focused on the questions how diffusive baryon transport manifests itself along the beam direction in hydrodynamic simulations and how it is affected by critical scaling of transport coefficients ($\tau_n$ and $\kappa_n$) and singularities in thermodynamic properties ($\chi$) in the proximity of the QCD CP. Based on the Hydro+/++ framework we identified the critical slowing down of the baryon diffusion current ($\tau_n\sim\xi^2$). The baryon diffusion flows observed in these simulations are characterized by an important feature: They show almost no sensitivity to critical effects even for fluid cells passing close to the CP. 

The main reasons for this insensitivity of baryon diffusion to critical dynamics are twofold: (i) The baryon diffusion flows are strong at early times but decay very quickly, before the system enters the critical region, because diffusion reduces the initially strong chemical gradients $\nabla(\mu/T)$ that drive it, and the baryon diffusion coefficient $\kappa_n$ describing the response to these gradients decreases quickly as the fireball cools by expansion. (ii) By the time the system reaches the phase transition, possibly passing close to the CP, the Navier-Stokes value of the baryon diffusion current is already very small; critical enhancement by the baryon diffusion coefficient $\kappa_n$ 
therefore does not help to revive it, and in any case the relaxation rate controlling the approach of $n^\mu$ to its critically affected Navier-Stokes value is reduced by critical slowing down. 

The observed insignificance of critical effects on baryon diffusion might be taken as permission to calibrate the fireball medium's bulk evolution at BES energies without worrying about critical effects on the baryon diffusion current. However, the possibility of critical effects on the shear and bulk viscous pressure evolution should also be kept in mind. Other aspects of the full dynamics may change the evolution of the temperature and chemical potential and thus induce sensitivity to the CP in the baryon sector as well. Only a full simulation including all dissipative effects simultaneously will allow us to quantitatively evaluate the significance of critical effects on the bulk medium evolution at BES energies.



\end{document}